\documentclass[preprint2]{emulateapj-rtx4}
\usepackage{rotating}
\usepackage{txfonts}
\usepackage{epsfig}
\usepackage{graphicx}
\usepackage{natbib} 
\usepackage{xspace}
\usepackage{color}
\usepackage[colorlinks,urlcolor=blue,citecolor=blue,linkcolor=blue]{hyperref} 

\begin{document}

\title{3.3 cm JVLA observations of transitional disks: \\  searching for centimeter pebbles}

\author{Luis A. Zapata\altaffilmark{1}, Luis F. Rodr\'\i guez\altaffilmark{1}, and Aina Palau\altaffilmark{1}}

\altaffiltext{1}{Instituto de Radioastronom\'\i a y Astrof\'\i sica, 
UNAM, Apdo. Postal 3-72 (Xangari), 58089 Morelia, Michoac\'an, M\'exico}

\email{lzapata@crya.unam.mx}
 
\begin{abstract}
We present sensitive (rms-noises $\sim$ 4 -- 25 $\mu$Jy) and high angular resolution ($\sim$1--2$''$) 8.9 GHz (3.3 cm) 
Karl G. Jansky Very Large Array (JVLA) radio continuum observations of 10 presumed transitional disks associated with 
young low-mass stars.  We report the detection of radio continuum emission 
in 5 out of the 10 objects (RXJ1615, UX Tau A, LkCa15, RXJ1633, and SR24s).  
In the case of LkCa15, the centimeter emission is extended, and has a similar morphology to that 
of the transitional disk observed at mm wavelengths with an inner depression. 
 For these five detections, we construct the Spectral Energy Distributions 
(SEDs) from the centimeter to submillimeter wavelengths, and find that they can be well fitted with a single (RXJ 1633 and UX Tau A) or 
a two component power-law (LkCa 15, RXJ 1615, and SR24s). For the cases where a single power-law fits well the data, the centimeter emission is 
likely produced by optically thin dust with large grains ({\it i.e.} centimeter-size pebbles) present  
in the transitional disks. For the cases where a double power-law fits the data, the centimeter emission might be produced by the combination 
of photoevaporation and/or a free-free jet. We conclude that RXJ1633 and UX Tau A are excellent examples of transitional disks where the structure of the 
emission from centimeter/millimeter pebbles can be studied. In the other cases, some other physical emitting mechanisms are also important 
in the centimeter regime.   
\end{abstract}  


\section{Introduction}

One of the first steps in the process of planet formation is that of dust grain must growth. The dust grains in a protoplanetary disk are expected to grow 
from sub-micron size (as observed in the Interstellar Medium, ISM) to millimeter/centimeter sizes, and then to large pebbles. The best evidence of such change comes 
from the relatively shallow wavelength dependence of the millimeter/centimeter thermal emission from dust \citep[e.g. CY Tau, CQ Tau, and DoAr 25:][]{tes2003,per2015}.
If we assume that the dust emission is optically thin,  then the estimated spectral index $\alpha$ (of the form S$_\nu$ $\propto$ $\nu^\alpha$, where S$_\nu$ is the
flux density and $\nu$ is the frequency) is directly related to the dependence of the opacity on the wavelength. Thus, the grain size can be deduced 
by the opacity law, but taking into account the optical depth and geometrical effects \citep[][]{tes2003}.  A clear example of the application of this technique 
has been done by \citet{wil2005}.
Using self-consistent disk models and deep VLA observations at 3.5 cm wavelengths towards the transitional disk of the T Tauri star TW Hya, \citet{wil2005} 
reported that a large amount of the orbiting particles in such disk should have agglomerated to centimeter and millimeter sizes providing some 
evidence at the beginning of the planet formation process.

The transitional disks were first indirectly inferred from their infrared spectra obtained by the {\it Infrared Astronomical Satellite (IRAS) and Spitzer Space Telescope}  
\citep{strom1989,cal2002, dal2005, kim2013}. This new class of disks are (proto)planetary disks around young stars  
which are optically thick and gas-rich, but which have AU-scale radial gaps or central depressions in their dust distribution. 
This gap is revealed as a deficit in the T Tau and Herbig Ae/Be star's infrared excess \citep[e.g.][]{kim2013, van2016b}, in more recently spatially resolved 
images of reflected light in the optical, and infrared \citep[{i.e.}][]{tal2010,can2013,may2012,ave2014}, and in the (sub)millimeter regime 
\citep{and2007, and2011,pie2006,ise2014,cie2012,oso2014}.  

However, the nature of the cavity has remained in dispute.  Up-to-date, there are many possibilities to explain its nature, 
some of them include:  photoevaporative winds,  dust  size  evolution,  and  tidal  interactions  with  stellar
or  planetary  companions \citep{will2011}.  Moreover, there is clearly molecular 
gas inside of these cavities, see for example: \citet{can2015,van2016,tang2016}, which 
indicates that such cavities are not really fully empty.

\begin{deluxetable*}{l c c c c c c c }
\tablecolumns{9} 
\tablewidth{0pc} 
\tablecaption{Parameters of the radio observations}
\tablehead{
\colhead{}                       &
\colhead{}                       &
\colhead{}                       &
\colhead{}                       &
\colhead{Rms}                    &
\multicolumn{2}{c}{Synthesized Beam} \\ 
\cline{6-7}
\colhead{}                       &
\multicolumn{2}{c}{Phase Center$^{a}$} &
\colhead{Phase}             &
\colhead{Noise}        &
\colhead{ Mayor $\times$ Minor axis}         &
\colhead{P.A.}                \\
\cline{2-3} 
\colhead{Source} & 
\colhead{$\alpha_{2000}$}  &
\colhead{$\delta_{2000}$}  &
\colhead{Calibrator} &
\colhead{[$\mu$Jy]}  &
\colhead{[arcsec $\times$ arcsec ]}   &
\colhead{[deg.] } 
} 
\startdata
UX Tau   A  & 04 30 04.00 & $+$18 13 49.3   & J0409$+$1217 & 4.1 & 1.03$\times$0.79 & $-$64 \\
DM Tau    & 04 33 48.74  & $+$18 10 09.7  & J0409$+$1217 & 5.1 & 1.00$\times$0.81 & $-$63 \\
LkCa15   &  04 39 17.79 & $+$22 21 03.2  & J0510$+$1800 & 3.5  & 0.98$\times$0.79 & $-$69 \\
SAO 2064 & 15 15 48.43  & $-$37 09 16.3   & J1454$-$3747 & 25$^{b}$  & 2.59$\times$0.72 &  $-$9\\
RXJ 1615  & 16 15 20.20  & $-$32 55 05.1   & J1607$-$3331 & 7.5 & 2.38$\times$0.73 &  $-$19\\
SR 24s      & 16 26 58.51 & $-$24 45 37.0   & J1626$-$2951  & 5.2 & 1.44$\times$0.76 & $+$8 \\
SR21        & 16 27 10.28 & $-$24 19 12.8  & J1626$-$2951  & 6.2 & 1.43$\times$0.74 & $+$4 \\
WSB 60     & 16 28 16.51  &$-$24 36 58.3   & J1626$-$2951  & 6.5 & 1.48$\times$0.73 & $+$6 \\
DOAR 44  & 16  31 33.46 & $-$24 27 37.4 	  & J1626$-$2951  & 12.6 & 1.27$\times$0.49 & $-$16 \\
RXJ 1633  & 16 33 55.60 & $-$24 42 05.0   &  J1626$-$2951  & 7.4 & 1.60$\times$0.79 & $-$14 \\

\enddata
\tablecomments{ (a) -- Units of right ascension are hours, minutes, and seconds,  and units of declination are degrees,
                           arcminutes, and arcseconds. The phase centers of these observations were obtained from \citet{and2011}, 
                           with the exception of that of RXJ 1633, which is from the SIMBAD database.\\
                           (b) -- The large rms-noise in the field of SAO 2064 is probably due to its southern declination, poor weather or
                                    a strong double source located within our FWHM.  The combination of all possibilities is also valid. }
\end{deluxetable*}

\begin{deluxetable*}{l c c c c c c c c c}
\tablecaption{Physical parameters of the transitional disks}
\tablehead{                        
\colhead{}                        &
\multicolumn{2}{c}{Position}  &
\colhead{}                              &
\colhead{}                              &
\multicolumn{4}{c}{Stellar and Accretion Properties} &   \\ 
\cline{2-3}
\cline{6-9}  
\colhead{}   &
\colhead{$\alpha_{2000}$}          &
\colhead{$\delta_{2000}$}           &
\colhead{Peak Flux}           &
\colhead{Flux Density }       &                            
\colhead{SpT}  &
\colhead{M$_*$}  &
\colhead{d}  &
\colhead{\.M} &  \\
\colhead{Source}                      &
\colhead{(h m s) }                     &
\colhead{($^\circ$ $^{\prime}$  $^{\prime\prime}$)}           &
\colhead{($\mu$Jy Beam$^{-1}$)}  & 
\colhead{($\mu$Jy)}  & 
\colhead{}  &
\colhead{(M$_\odot$)}  &
\colhead{(pc)} &
\colhead{(M$_\odot$ yr$^{-1}$)} & 
}
\startdata
UX Tau  A  &   04 30 03.996 & $+$18 13 49.20  &   55  $\pm$ 6  &  57  $\pm$ 8  & G8    & 1.5    &      140       & 10$^{-8}$ \\
DM Tau     &   04 33 48.742 & $+$18 10 09.74   &   $\leq$20.4    &   -                  &  M1   & 0.53   &     140       & 6 $\times$ 10$^{-9}$\\  
LkCa15     &   04 39 17.794 & $+$22 21 03.29   &  31 $\pm$ 15 &  45 $\pm$ 17 &  K3    & 1.01  &      140      &  2 $\times$ 10$^{-9}$\\ 
SAO 2064 &   15 15 48.447 & $-$37 09 16.42    &  $\leq$100 &   -                       &  F4    & 1.6    &      140      &  6 $\times$ 10$^{-9}$   \\
RXJ 1615  &   16 15 20.238 & $-$32 55  05.96   &  45 $\pm$ 8  &   62 $\pm$ 15 &  K5    &  1.1    &      185     &   4 $\times$ 10$^{-10}$\\
SR 24s        &  16 26 58.502 & $-$24 45  37.25   &  90 $\pm$ 6  &   96 $\pm$ 7 &   K2  &  2.0    &      125     &  10$^{-8}$\\
SR 21        &   16  27 10.265  & $-$24 19 13.24   &   $\leq$25.6  & -                    &   G3   &  2.0    &       125     & 2 $\times$ 10$^{-9}$ \\ 
WSB 60    &   16 28 16.532 & $-$24 36 58.45    &  $\leq$26 &   -                         &   M4   &  0.25   &      125     & 2 $\times$ 10$^{-9}$\\ 
DOAR 44  &   16 31 33.458 & $-$24 27  37.33   &     $\leq$50.4 &   -                   &   K3  & 1.3       &       125     &  $<$ 10$^{-9}$ \\   
RXJ 1633    &  16 33 55.611 & $-$24 42  05.43   & 216 $\pm$ 10   &  218 $\pm$ 9 &  K7 & 0.7    & 120 & 10$^{-10}$ \\
\enddata
\tablecomments{These values were obtained from the task IMFIT of CASA. The upper limits presented here are 4$\sigma$. \\
                           The stellar and accretion properties of these disks were obtained from \citet{and2011} and \citet{cie2012}. }
\end{deluxetable*}

Optical and radio observations have revealed that the transitional and pre-transitional disks \citep[for a more complete 
review about this kind of disks, see:][]{espa2014} have associated ionized jets, 
and can be detected as weak free-free sources \citep{eis2000, rod2014, mac2016}. In the case of AB Aur, a well studied transitional disk, 
\citet{rod2014} concluded that the radio centimeter emission is tracing a collimated and ionized outflow, given the morphology, 
orientation, spectral index, and lack of temporal variability of the centimeter source. This case supports the interpretation that 
probably the radio emission at these wavelengths, and from this class of objects, could be arising from a faint free-free jet that 
is still present at the most evolved phases of the formation of a star. 
AB Aur is a Herbig Ae star with a mass of 2.4 $\pm$ 0.2 
M$_\odot$, a total luminosity of about 38 L$_\odot$, a spectral type A0, and an estimated age of 4 $\pm$ 1 Myr  \citep{dew2003}. 
These properties make to AB Aur certainly a bigger and hotter young star hosting a transitional disk compared with the group 
of stars analyzed here, which are more typically associated with colder solar-type objects in probably a more evolved phase, see  \citet{and2007}. 
AB Aur is also surrounded by a large envelope \citep{tang2016}. However, one could argue that in 
the case of GM Aur, a transitional disk that is associated with a T-Tauri young star 
($d\simeq$140 pc, K5 spectral type, $L_{\star}\simeq0.9$ $L_{\odot}$, $M_{\star}\simeq1.1~M_{\odot}$; \citealp{ken95}),  \citet{mac2016} reported a compact free-free jet.   
GM Aur does not have a large accreting envelope as in the case of AB Aur \citep{mac2016}.  Hence, one might conclude that the radio jets are present even
when the transitional disks are not relatively young. 

We  report here on deep JVLA 3.3 centimeter observations of ten presumed  transitional disks associated with 
young low-mass stars.  We detected radio emission in only 5 out of the 10 observed disks.  The radio emission 
is mostly unresolved and very faint. However, in the case of LkCa 15 the emission is well resolved. 
The nature of the radio emission present in the disks is discussed in the following sections.

\begin{figure*}
\centering
\includegraphics[angle=0,scale=0.45]{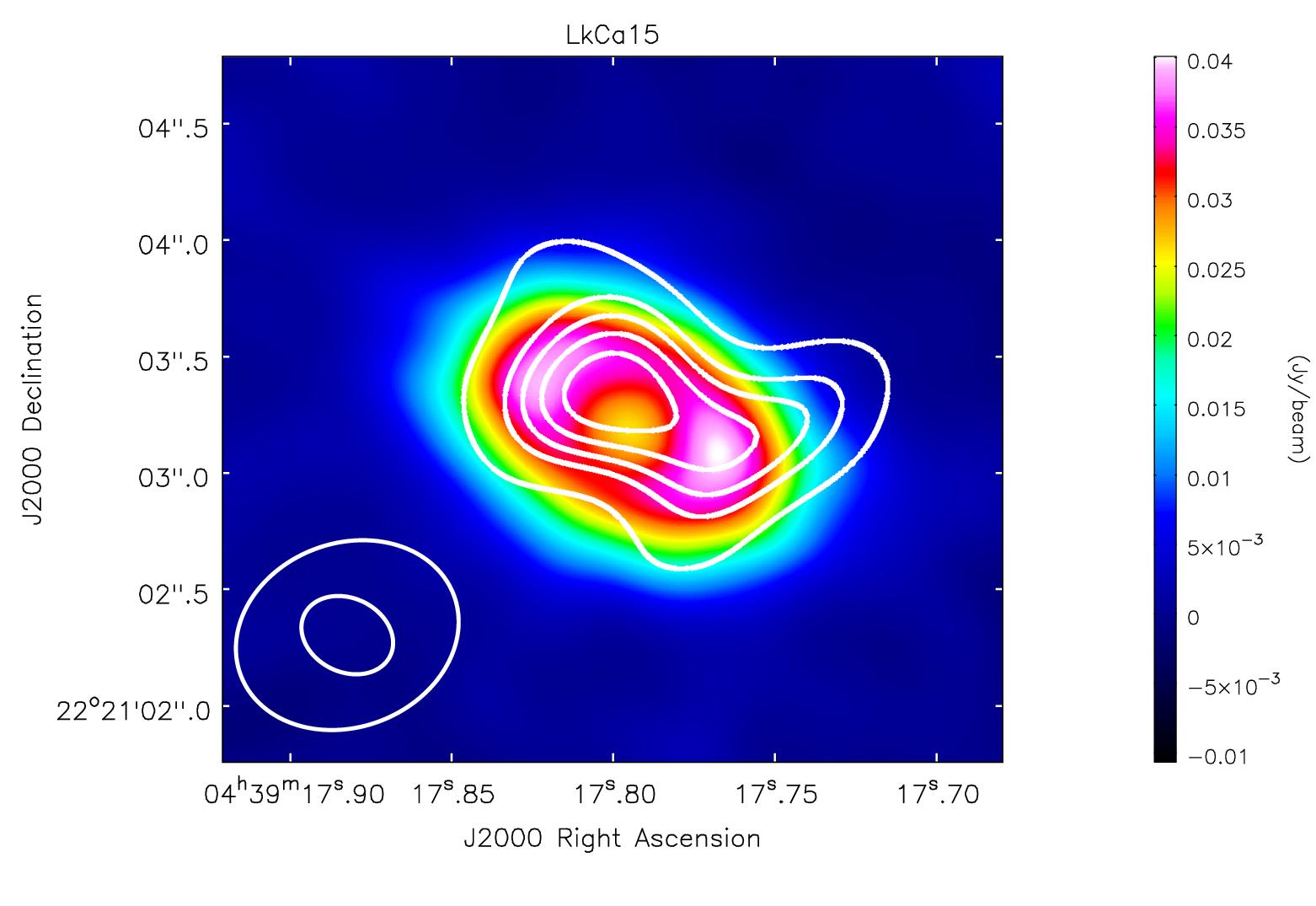}
\includegraphics[angle=0,scale=0.45]{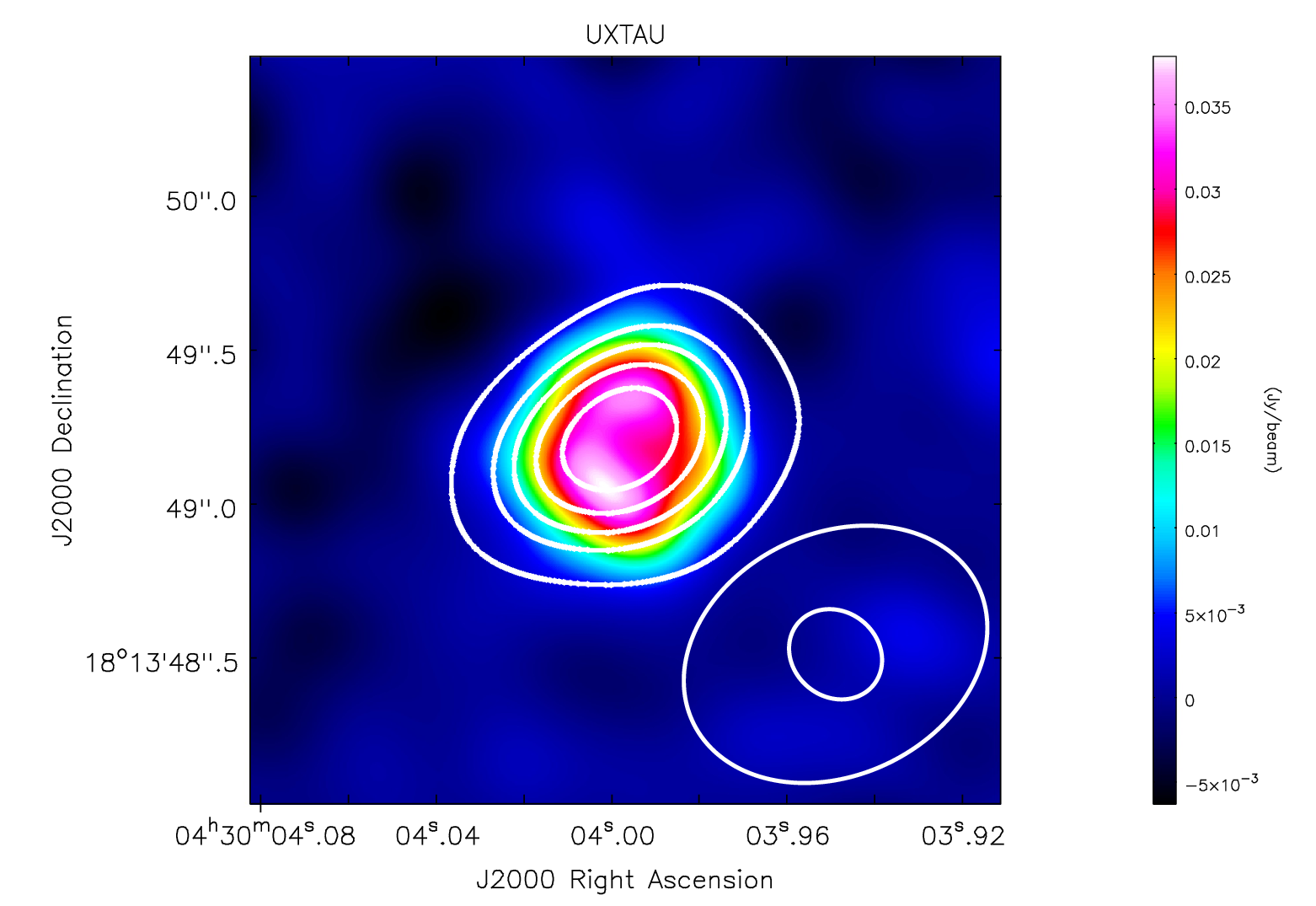}
\includegraphics[angle=0,scale=0.45]{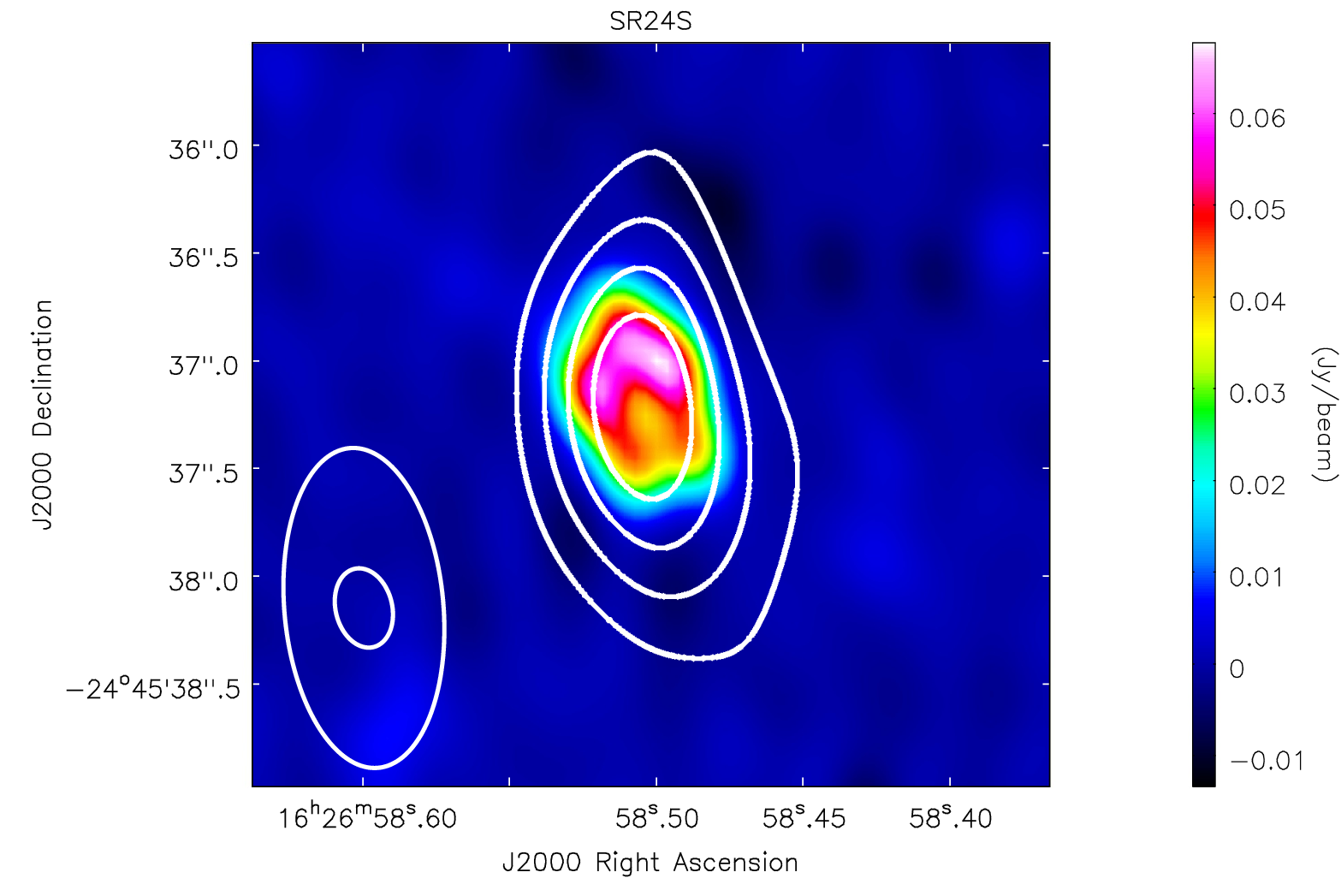}
\includegraphics[angle=0,scale=0.45]{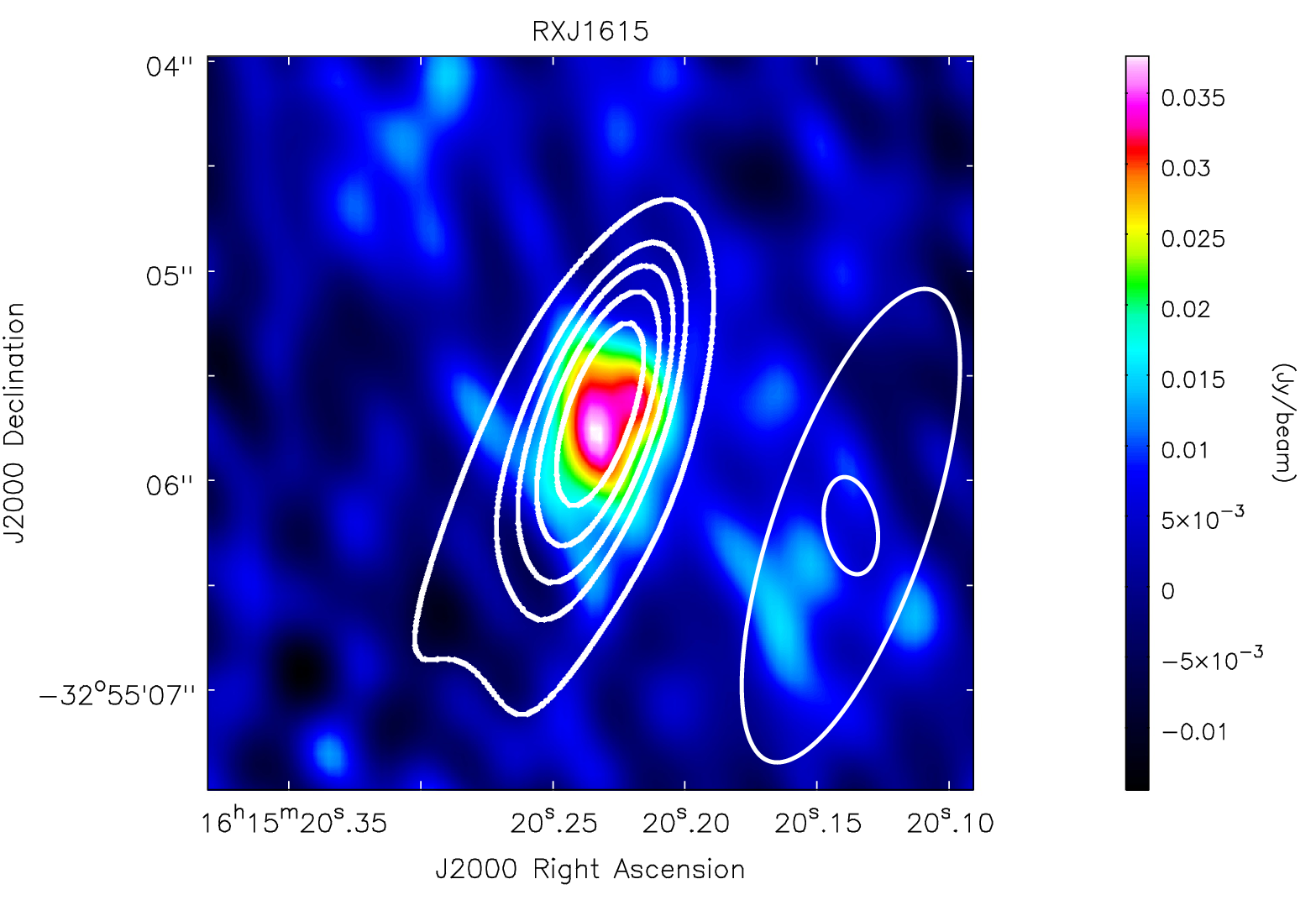}
\includegraphics[angle=0,scale=0.45]{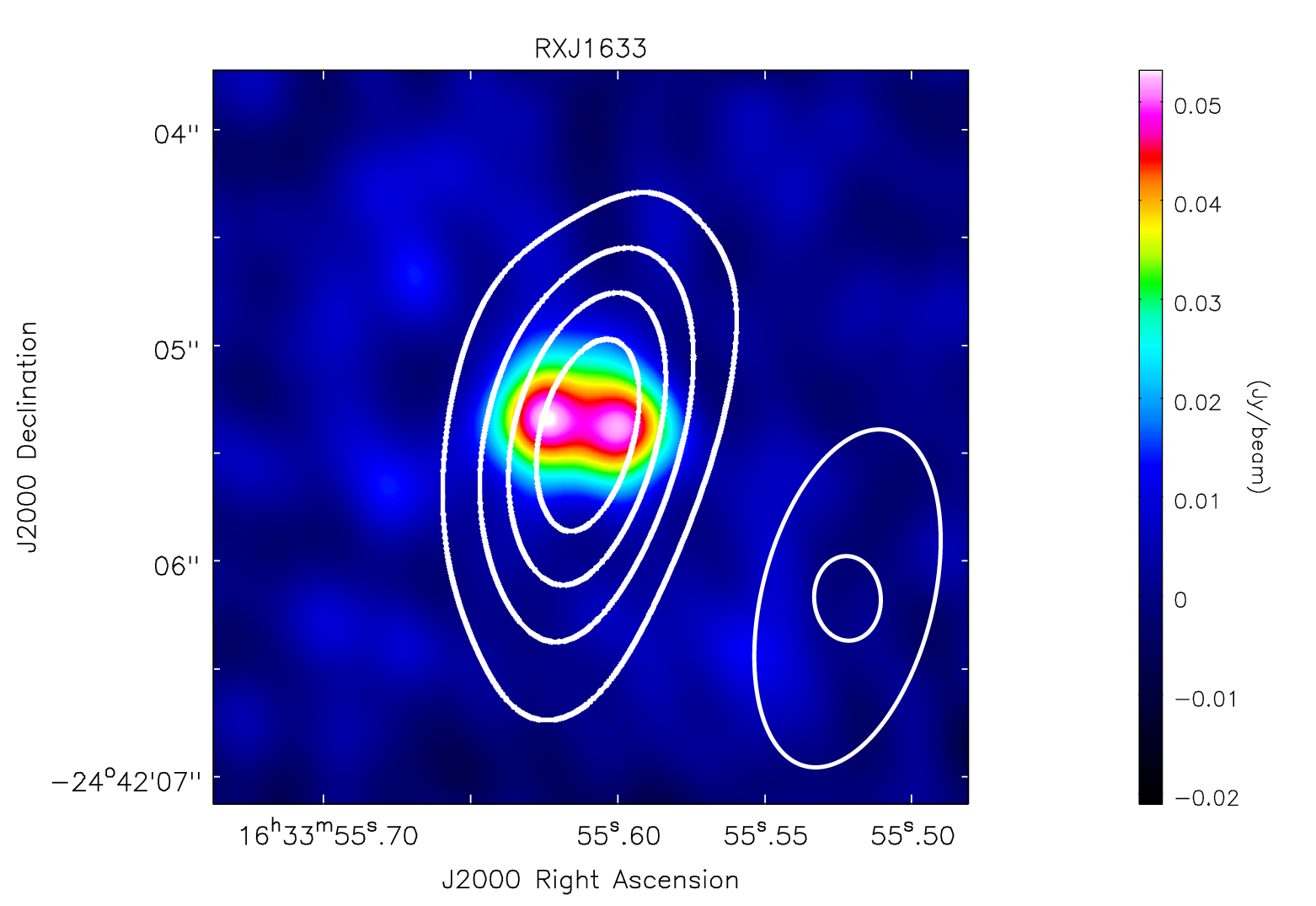}
\caption{VLA 3.3 cm continuum emission from the five detected transitional disks.  
 The centimeter emission is shown in white contours overlaid on a colour image of the 0.88 mm emission from \citet{and2011}. 
 For LkCa 15, the contours are 2.8, 4.8, 5.7, 6.8, and 8.0 $\times$ $\sigma$, the rms-noise of the image. 
 For UX Tau A, the contours are 3.6, 6.6, 8.0, 9.5 and 10.2 $\times$ $\sigma$, the rms-noise of the image.
 For  SR 24s, the contours are  3.6, 7.3, 11.1, and 15.1 $\times$ $\sigma$, the rms-noise of the image.
 For RXJ 1615, the contours are  2.8, 3.7, 4.4, 5.0 and 6.0 $\times$ $\sigma$, the rms-noise of the image.
 For RXJ 1633, the contours are 6.0, 11.8, 17.8, and 24.0 $\times$ $\sigma$, the rms-noise of the image.
 The half-power contour of the synthesized beam of the 3.3 cm emission is shown as the large ellipse in the bottom 
  left or right corners of each image. 
 The small ellipse is the half-power contour of the synthesized beam of the 0.88 mm emission also in each image.}
\label{fig1}
\end{figure*}

\section{Observations}

The observations were carried out with the Karl G. Jansky Very Large Array of NRAO\footnote{The National 
Radio Astronomy Observatory is a facility of the National Science Foundation operated
under cooperative agreement by Associated Universities, Inc.} centered at a rest frequency of  8.9 GHz (3.3 cm) during
2013 October and 2014 January.  At that time the array was in its  B  configuration.  We used 26 antennas of the array, yielding 
baselines with projected lengths from 10 to 350 k$\lambda$. 

The absolute amplitude calibrators were J1331+3030 and J0137+3309, while the gain-phase calibrators are presented in Table 1. 
The integration time in each source shown in Table 1 was about 20 min. The weather conditions were good and stable, 
with an average precipitable water vapour of about 7.0 mm.

The digital correlator of the JVLA was configured in 16 spectral windows of 128 MHz width divided 
in 64 channels. The total bandwidth for the observations was 2.048 GHz in a dual-polarization mode.

The data were analyzed in the standard manner using the CASA (Common Astronomy Software Applications) package of NRAO. 
We tried to apply self-calibration to the data, however,
given that all sources are very faint objects ($\ll$ 1 mJy), we were not successful in improving the images.
We used a natural weighting in order to obtain a slightly better sensitivity losing some angular resolution. 
To construct the continuum we used a bandwidth of 2.048 GHz.
The resulting image {\it rms}-noises and angular resolutions are presented in Table 1 and 2.  The resulting continuum images were
made using CASA with the task CLEAN with mode mfs \citep{hog1974}. The images obtained at these centimeter wavelengths were overlaid on the Submillimeter Array 
images obtained by \citet{and2011}. The uncertainty in the flux scale is estimated to be between 5\% and 10\%.

In this paper, we have additionally used centimeter and millimeter data from the VLA archive. 
Following the equation A5 of \citet{pal2010}, we have estimated the Largest Angular Scales (LAS) that interferometers are sensitive 
for the JVLA 7 mm and the SMA 0.85 mm observations, see Table 3. We obtained a LAS for archive 7 mm of 21.2$''$, while for the SMA 0.85 mm observations
a LAS of 4.8$''$ is estimated. As the LAS of the 7 mm archive observations is about of a factor of 4 larger as compared to that of 0.85 mm, this indicates that we are
are not filtering out much emission at 7 mm, and the flux densities at this wavelength presented in Table 3 can be well used  to construct the SEDs
for the objects LkCa 15, UX Tau A, and SR 24 S.        

\begin{figure}
\centering
\includegraphics[angle=0,scale=0.46]{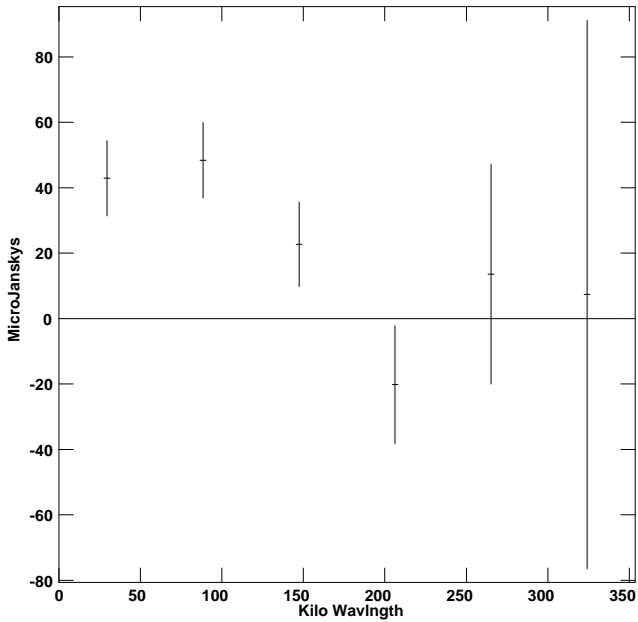}
\caption{Azimuthally averaged and radially binned visibility profile plotted in $\sim$70 kilo$\lambda$ bins of  LKCa 15 transitional disk.
               Note that even when there are large errors bars in our measurements, a depression about 200 k$\lambda$ 
               appears present, which suggests the presence of an inner hole in the disk.  }
\label{fig2}
\end{figure}
 
\section{Results}

We have observed 10 transitional disks at a wavelength of 3.3 cm using the Karl G. Jansky Very Large Array (JVLA) as a continuation of 
our study to better understand the nature of the centimeter emission in these systems \citep{rod2014}. We detected compact emission 
in only five transitional disks: RXJ 1615, UX Tau A, LkCa15, RXJ 1633, and SR 24s. 
 We looked for a specific trend (in the mass accretion rates, ages, and even spectral types) to explain the absence of radio emission in the 
rest of the transitional disks (DM Tau, SAO 2026, SR21, WSB60, and DOAR 44), unfortunately a trend is not obvious, see Table 2. 

In Table 2, we give the physical parameters of the five detected disks at these radio wavelengths 
as well as the upper limits for the non-detections.  
We detected these sources at more than a 4$\sigma$-level, equivalent to a range of 20-100 $\mu$Jy.
The radio emission arising from the five sources is mostly unresolved, with LkCa15 being the only source resolved.  

The centimeter emission from LkCa15 is extended, and has a similar morphology (2.6$''$ $\pm$ 0.1$''$ $\times$ 1.4$''$ $\pm$ 0.2$''$; PA: 67$^\circ$ $\pm$ 2.0$^\circ$;
this corresponds to spatial scales of 364 AU $\times$ 196 AU) 
to that of its associated mm transitional disk \citep[1.3$''$ $\pm$ 0.06$''$ $\times$ 0.8$''$ $\pm$ 0.04$''$; PA: 65$^\circ$ $\pm$ 4.4$^\circ$;][]{and2011} 
with an inner depression, see Figure 1. We have plotted the real component of the visibility profile (Figure 2), which shows the characteristic null 
and negative region that confirm the central hole in the disk, see for example \citet{oso2016}. The null falls around 200 k$\lambda$, 
corresponding to hole radii of about 70 AU.  The centimeter radio emission is resolved, and shows 
a similar morphology to the disk. This therefore suggests that the 3.3 cm radio emission
from LkCa15 is originating from the transitional disk instead of, for example, a thermal free-free jet, see \citet{rod2014}. 
We discuss in detail the nature of the cm emission in LkCa15 below.

\begin{deluxetable}{l c c c c}
\tabletypesize{\scriptsize}
\tablecaption{Submillimeter to centimeter flux densities}
\tablehead{                        
\colhead{Source}              &
\colhead{$\lambda$/Freq. }     &
\colhead{Flux Density }    &      
\colhead{Reference}        &  \\
\colhead{ }              &
\colhead{[cm/GHz]}        &
\colhead{[mJy]}      &      
\colhead{}               &
}
\startdata
RXJ16      &               &      &     \\  
                     &  0.088/340    &  430 $\pm$ 30           &   \citet{and2011}   \\  
                     &  0.140/214    &  169 $\pm$ 10           &     \citet{uba2012} \\ 
                     &  0.318/94    &   6.8 $\pm$ 1.2          &     \citet{uba2012} \\ 
                     &  0.700/43    &   0.8 $\pm$ 0.2          &     \citet{uba2012} \\ 
                     &  3.300/8.9    &   0.062 $\pm$ 0.015    &    This paper \\
                     &      &      &    \\
UXTauA    &      &      &    \\
	             &  0.045/666    &  523 $\pm$ 37                  &   \citet{and2005}   \\ 
                     &  0.088/340     &  173 $\pm$ 3                    &   \citet{and2005}   \\ 
                     &  0.130/230    &    52 $\pm$ 2                    &   \citet{jen2003}   \\
                     &  0.300/100    &    10.1 $\pm$ 0.3              &   \citet{pin2014}   \\
                     &  0.710/44    &    1.01 $\pm$ 0.06              &   VLA Archive   \\
                     &  0.920/32    &    0.47 $\pm$ 0.02              &   VLA Archive   \\
                     &  3.300/8.9    &    0.057 $\pm$ 0.008        &   This paper  \\
                     &      &      &    \\
LkCa15       &      &      &   \\
                    &  0.088/340   &  430 $\pm$ 30           &   \citet{and2005}   \\ 
                    &  0.130/230    &  167 $\pm$ 6           &   \citet{and2005}   \\  
                    &  0.280/107    &  17 $\pm$ 0.8           &   \citet{pie2006}   \\ 
                    &  0.700/43    &  0.60$\pm$ 0.10           &  VLA Archive    \\
                    &  3.300/8.9    &  0.045 $\pm$ 0.017        &   This paper   \\
                    &  6.000/5.0    &  $\leq$ 0.021         &   VLA Archive   \\
                    
                      &      &      &    \\
RXJ1633       &      &      &   \\
                      &  0.088/340  &  2200 $\pm$ 300           &   \citet{cie2012}   \\
                      &  0.130/230    &  818 $\pm$ 100               &   \citet{cie2012}   \\ 
                      &  3.300/8.9    &   0.218 $\pm$ 0.009      &    This paper. \\
SR24s           &      &      &   \\
                      &      &      &    \\
                     &  0.088/340    &  430 $\pm$ 30           &   \citet{and2007}   \\ 
                     &  0.130/230    &  197 $\pm$ 20           &   \citet{ise2009}   \\ 
                     &  0.300/100    &   26.6 $\pm$ 1.5          &     \citet{ric2010} \\ 
                     &  0.730/44    &   1.03 $\pm$ 0.18         &   VLA Archive  \\ 
                     &  0.880/34    &   0.67 $\pm$ 0.03         &   VLA Archive  \\
                     &  3.300/8.9    &   0.096 $\pm$ 0.004      &   This paper \\
ABAur          &      &      &   \\
                     &      &      &    \\ 
                     &  0.075/400    &  530 $\pm$ 90           &   \citet{ack2004}   \\
                     &  0.085/352   &  360 $\pm$ 70           &   \citet{ack2004}   \\
                     &  0.110/272    &  150 $\pm$ 20           &   \citet{ack2004}   \\                   
                     &  0.270/111    &  11 $\pm$ 0.5           &   \citet{ack2004}   \\
                     &  0.700/43    &  0.85 $\pm$ 0.03           &   \citet{rod2014}   \\
                     &  3.300/8.9    &  0.13 $\pm$ 0.02           &   \citet{rod2014}   \\
                     &  4.500/6.6    &  0.085 $\pm$ 0.01           &   \citet{rod2014}   \\
\enddata
\end{deluxetable}

\begin{figure*}
\centering
\includegraphics[angle=0,scale=0.94]{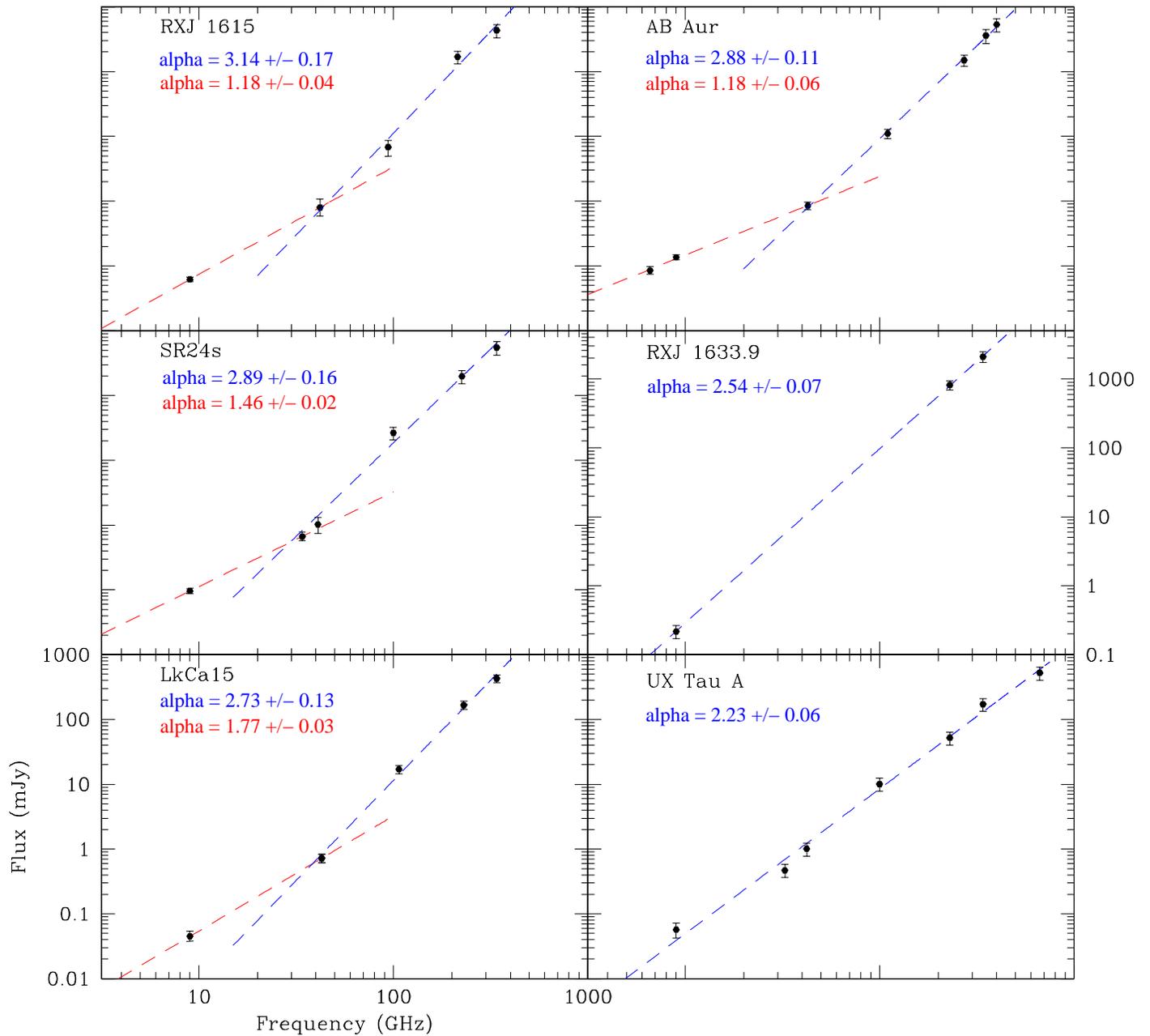}
\caption{Spectral Energy Distributions (SEDs) of the detected transitional disks from the centimeter to (sub)millimeter wavelengths. 
              The dashed lines are a least-squares power-law fit (of the form S$_\nu$ $\propto$ $\nu^\alpha$) to the spectrum. 
              The $\alpha$-values of the fitting for the different components of the spectrum are shown in the panels. The data are presented
              in Table 3. The axis scales are the same for all objects, but only in the case of RXJ 1633 
              the scale in y-axis is different. }
\label{fig3}
\end{figure*}

For the rest of the disks: RXJ 1615, UX Tau A, RXJ1633, and SR 24s, the 3.3 cm radio emission is unresolved, see Figure 1.   
However, the radio emission is well centered at the position of the central young star, see Table 1 and 2. None of these objects shows a central 
depression, as in the case of LkCa15. This is probably due to the poor angular resolution of our radio observations as compared to that 
obtained with the Submillimeter Array observations \citep{and2011}. There is a factor of about four between both angular resolutions. 
Deep 3.3 cm radio observations with the JVLA in the A-configuration will help in resolving the emission at these wavelengths from 
the transitional disks, perhaps revealing the inner holes, as those mapped at submillimeter wavelengths.     

In Table 3 we give the values of the flux densities from submillimeter to centimeter wavelengths of the five detected transitional disks
that were compiled from the literature, the JVLA archive, and from our centimeter observations.   From these data, we constructed 
the Spectral Energy Distributions (SEDs) from the submillimeter to centimeter regimes, and found that they can be well fitted 
with a  double or single power-laws.  
In order to fit the data, the error bars were calculated adding in quadrature systematic
errors of order of 5-25\% (which are typical observational flux uncertainties) to the formal fluxes uncertainties 
(obtained from a Gaussian fitting in our data) to obtain a reduced  $\tilde\chi^2=1$. We used a least-squares fitting routine to obtain these results.
If a single power law failed to fit all data with the introduced errors of around 5-25\%, we then tried to use a two-component law to fit the centimeter 
and (sub)millimeter data separately. Here, we consider the centimeter regimen to have wavelengths $>$ 1 cm.

The best cases for this single power-law fitting are UX Tau A and RXJ 1633. However, for the latter case, we only found three flux 
measurements at these wavelengths, see Table 2, and Figure 3.  For the cases of LkCa15, RXJ 1615, and SR 24s, these can be 
fitted by a two component power-law, see Figure 3.  These two power-laws fit the centimeter and submillimeter regimes.

\section{Discussion}

In Figure 3, we have additionally included the SED of AB Aur 
(a transitional disk with a free-free jet) at these similar wavelengths \citep{rod2014}, see Figure 3. 
The SED of AB Aur has a two-component  spectrum in which the centimeter emission is dominated by a slowly rising spectrum ($\alpha$ $=$ 1.18), which can be interpreted as
moderately optically thick free-free emission (from a faint jet), while the millimeter and submillimeter emission is dominated by a component
that rises rapidly with frequency ($\alpha$ $=$ 2.88),  which is interpreted as thermal emission from optically thin dust, see Figure 3. 
This spectrum is very similar to that observed in LkCa15, RXJ 1615, and SR 24s. 
This suggests that the centimeter emission from these three sources might be arising from a thermal free-free jet. However, as this radio emission is resolved
in the case of LkCa15, and has a similar morphology to that of the transitional disk, then we propose that more likely the radio emission could come from 
a photo-evaporatated disk in this case.  Photoevaporation, together with viscous accretion, is expected to play an important role in the dispersal of protoplanetary disks
\citep{mac2016}. High energy radiation -- i.e. UV and X-ray radiation -- originating at the stellar chromosphere of low-mass stars
 can ionize and heat the disk surface forming thus a phoevaporative wind \citep{gor2009}.
For the cases of RXJ 1615, and SR 24s, their radio emission is unresolved and it is not possible to discard if a thermal jet is present or if the photoevaphoration is playing
an important role in producing the radio emission.

For the cases of UX Tau A and RXJ 1633, none of their spectra from the detected transitional disks has such a behaviour and they can be well fitted by a single component.  Such a single 
component which rises rapidly with frequency is produced by optically thin dust emission with approximately a Rayleigh-Jeans exponent from the transitional disks \citep{pie2006, hug2009,and2011}. This is the case even at centimeter wavelengths for the two disks as our SEDs demonstrate. 
For pre-transitional or transitional disks it is expected to find inner holes at similar wavelengths.  However, at the submillimeter regime, 
it is expected that the innermost parts of younger disks will be optically thick, see for example the case of the disk IRAS 16293$-$2422B,  \citet{zap2013}.

As the 3.3 cm continuum emission is arising from optically thin dust from the transitional disks in the cases of UX Tau A and RXJ 1633 --as it is shown by their SEDs at centimeter 
and submillimeter wavelengths, we then suggest that such a centimeter emission detected here is arising from the disk instead from a thermal free-free jet. 
The dust grains contained in the disks must then be larger than the ISM grains, probably with sizes of the order of some centimeters.  
The $\alpha$-values of 2.22 and 2.54 found in their SEDs imply a $\beta$ (the power-law spectral index of the
dust opacity with the form: $\kappa_\nu=\nu^\beta$, where $\kappa$ is the dust opacity coefficient and $\nu$ is the frequency) 
less than the unity, so grain growth should occur \citep{per2015,pin2014,uba2012,ric2010}. We would like to note again that as the emission from the
transitional disks is mostly optically thin, then  $\beta \leq 1$ might imply grain growth, see \citet{can2015,dra2006,van2013,pin2015}.   
This physical phenomenon is expected as the transitional disks are protoplanetary disks in an evolved stage, where the planet 
formation is supposedly occurring \citep{kim2013}. However, we note that in addition to optically
thick clumps in the disk, there are several factors that can influence $\beta$, as e.g., porosity or grain size distribution. 
While ISM grains are characterized by a grain size distribution following a power-law with index
$p = 3.5$, the size distribution when grains grow up to large sizes is not well constrained, see for a review: \citet{dra2006}.
We also note that the $\alpha$-values determined in this study for the (sub)millimeter regimen and shown in Figure 3 are in well agreement 
to the mean value reported in \citet{pin2014} for a group of transitional disks.  
  
Further sensitive observations with a high angular resolution at these centimeter wavelengths might confirm the presence 
of two families of transitional disks. On one hand there are disks like the ones reported in this study and in others e.g. \citet{per2015}  
(UX Tau A, RXJ 1633, and TW Hya, and CQ Tau) where the centimeter emission is 
arising from large grains located in the disks, and without the presence of a thermal radio jet or strong photoevaphoration. 
On the other hand, objects like AB Aur, LkCa 15,  RX J1615, and GM Aur with a transitional 
disk and with the presence of a faint ionized jet or photoevaphoration. The SEDs and this kind of centimeter observations will help in discriminating between 
both cases. However, we speculate that these two populations of transitional disks might be explained 
as part of an evolutionary trend, where younger transitional disks still would have important accretion, generating the presence of ionized jets, i.e. AB and GM Aur-like objects. 
More cases where ionized jets are (or not) present in transitional disks could help in a better understanding of this tentative hypothesis.

\section{Conclusions}

The high sensitivity and angular resolution of the Karl G. Jansky Very Large Array observations allowed 
to detect some of the classical transitional disks at these radio wavelengths. 
The main results of our work can be summarized as follows.

\begin{itemize}

\item We report the detection of unresolved (with the exception of LkCa15) radio continuum emission in 5 out of the 10 disks.  
In the case of LkCa15, the radio emission is extended, and has a similar morphology to that of 
the mm transitional disk with an inner depression.  We suggest that the cm emission is tracing a photoevaporating disk. 
For the rest of the disks, the unresolved radio emission peaks close to the position of the young star.   

\item For these five detections, we construct the Spectral Energy Distributions 
(SEDs) from the centimeter to submillimeter wavelengths, and find that they can be well fitted with a single or with two power-laws, 
with $\alpha$-values something flat or very step.

\item Our results suggest that the emission detected at these centimeter wavelengths is likely produced 
by optically thin dust with large grains ({\it i.e.} centimeter-size pebbles) present in the transitional disks UX Tau A and RXJ 1633,
and from a thermal jet or a photoevaphorative wind in the cases of LkCa 15,  RX J1615 and SR24s.

\item We conclude that higher angular resolution JVLA observations, especially in the A configuration, and 
maybe using the X or Ka bands are needed to resolve the radio emission present in the other four 
transitional disks (RXJ 1615, UX Tau A, RXJ 1633, and SR 24s). For the case of LKCa 15, such observations will help
in better tracing its morphology. This might help in testing our hypothesis that the cm emission is arising from 
large dust grains in the disks in the cases of UX Tau A and RXJ 1633.

\item Our data suggest the possibility of having two populations of transitional disks, one with the presence 
of ionized jets and the other one without these jets. Radio images of a larger number of transitional 
disks may help in confirming this trend.

\end{itemize}

\acknowledgments
We thank Sean Andrews and Ezequiel Manzo for providing the SMA images of the transitional disks, and for the models 
for some disks presented in this work that guided our interpretation. 
We are very thankful for the thoughtful suggestions of the anonymous referee that helped to improve 
our manuscript significantly.
This research has made use of the SIMBAD database, operated at CDS, Strasbourg, France. 
L.A.Z., and L.F.R. are grateful to CONACyT, Mexico, and DGAPA, UNAM for their financial support.
A.P. acknowledges financial support from UNAM-DGAPA-PAPIIT IA102815 grant, M\'exico.

{\it Facilities:} \facility{The Karl G. Jansky Very Large Array (JVLA)}, \facility{The Submillimeter Array (SMA)}.

\end{document}